\documentclass[aps,reprint,prl,unsortedaddress,superscriptaddress,letterpaper]{revtex4-1}

\newcommand{\apjl}{Astrophys. J. Lett.}
\newcommand{\grl}{Geophys. Res. Lett.}

\newcommand{\aap}{Astron. Astrophys.}
\newcommand{\mnras}{MNRAS}

\newcommand{\rsi}{Rev. Sci. Instrum.}
\usepackage{graphicx}

\begin{document}

\title{Measurement of anomalously strong emission from the 1s-9p
  transition in the spectrum of H-like phosphorus following charge
  exchange with molecular hydrogen}

\author{M.~A.~Leutenegger} 

\affiliation{NASA Goddard Space Flight Center, Greenbelt, Maryland
  20771, USA}
\author{P.~Beiersdorfer} 
\affiliation{Lawrence Livermore National Laboratory, Livermore,
  California 94550, USA}
\affiliation{Space Sciences Laboratory, University of California,
  Berkeley, California 96720, USA} 
\author{G.~V.~Brown} 
\affiliation{Lawrence Livermore National Laboratory, Livermore,
  California 94550, USA}
\author{R.~L.~Kelley} 
\affiliation{NASA Goddard Space Flight Center, Greenbelt, Maryland
  20771, USA}
\author{C.~A.~Kilbourne}
\affiliation{NASA Goddard Space Flight Center, Greenbelt, Maryland
  20771, USA}
\author{F.~S.~Porter}
\affiliation{NASA Goddard Space Flight Center, Greenbelt, Maryland
  20771, USA}

\date{\today}

\begin{abstract}

We have measured K-shell x-ray spectra of highly ionized argon and
phosphorus following charge exchange with molecular hydrogen at low
collision energy in an electron beam ion trap using an x-ray
calorimeter array with $\sim$6 eV resolution. We find that the
emission at the high-end of the Lyman series is greater by a factor of
2 for phosphorus than for argon, even though the measurement was
performed concurrently and the atomic numbers are similar. This does
not agree with current theoretical models and deviates from the trend
observed in previous measurements.

\end{abstract}

\maketitle

Charge exchange (CX) between neutral species and ions is important in
setting the ionization balance in laboratory and astrophysical
plasmas, as well as the storage time in ion traps and storage rings,
antihydrogen production, and diagnosing fusion plasmas
\cite{1977PhRvL..38.1359I, 1982PhRvL..49..737F, 1984PhRvL..52..530F,
  1984PhRvL..52.2245K, 1986PhRvL..56...50R, 1998PhRvA..58.2043S,
  2001PhRvL..86..636G, 2007PhRvL..98k3002G}.  X rays from CX between
solar wind ions and neutrals in comets and planetary atmospheres are
highly diagnostic of both the solar wind and the neutral species
\cite{1997GeoRL..24..105C, 2002Natur.415.1000G, 2006A&A...451..709D,
  2007A&A...469.1183B}. CX between the solar wind and neutral
heliospheric gas contributes a variable background that needs to be
understood in order to interpret observations of interstellar and
intergalactic plasmas \cite{2000ApJ...532L.153C}. Solar system CX
typically occurs at low collision velocities ($< 700$ km/s), a
previously neglected regime that has been the subject of renewed
interest over the last decade \cite{2000ApJ...533L.175G,
  2007PhRvA..75c2704M, 2008CaJPh..86..151W}.

Qualitatively, one expects CX between given neutral and highly ionized
species to be dominated by capture of single electrons into states
with a particular principal quantum number around $n_c \approx
q^{3/4}$, where $q$ is the charge of the capturing ion
\cite{1981PhRvA..24.1726O}. Following capture, the electron decays
radiatively to the ground state, either directly or via a cascade. For
bare ions, captures into $n_c p$ usually decay directly to the $1s$
ground level, while captures into higher angular momentum states will
cascade on the Yrast chain ($l = n - 1$) in steps of $\Delta l, \Delta
n = -1$, ultimately resulting in a $2p \rightarrow 1s$ Ly$\alpha$
photon.

In the limit of high collision velocity, the angular momentum
distribution of captured electrons is expected to be statistical, and
high angular momentum states dominate; however, at low collision
velocity, high angular momentum states are not accessible
\cite{1985JPhB...18..737D, 1986JPhB...19L.507B}. Because captures into
$n_c p$ usually result in $n_c p \rightarrow 1s$ photons, while
captures into other states usually result in Ly$\alpha$ photons, the
ratio of flux in the $n_c p \rightarrow 1s$ transition to that in
Ly$\alpha$ is a diagnostic of the distribution of angular momentum
states populated, and thus of the collision velocity. X-ray detectors
with moderate spectral resolution do not resolve the high-$n$ Lyman
series; thus it is often simpler to measure the hardness ratio,
defined as ${\cal H} \equiv F_{3+} / F_2$, where $F_2$ is the
observed flux of transitions from principal quantum number $n = 2$ to
the ground state, and $F_{3+}$ is the sum of flux in transitions from
$n \ge 3$ to the ground state.

\citet{2000PhRvL..85.5090B} measured K-shell x-ray spectra from CX at
low ion temperature for a wide range of ions in the Lawrence Livermore
National Laboratory (LLNL) EBIT-I electron beam ion trap (EBIT) using
a high purity germanium detector, with the goal of studying the
angular momentum distribution of captured electrons in CX at low
collision velocity. Not only was the expected deviation from a
statistical population of angular momentum states found, but the
observed hardness ratios were much higher than predicted in classical
trajectory Monte Carlo (CTMC) theory, with a hardness ratio of about
unity observed over a wide range in atomic number (10-54) for H-like
species.

A subsequent experiment by the Berlin EBIT group has reproduced the
measurement of the LLNL EBIT group for argon; however, when they
instead extracted a beam from their EBIT and collided it with a
stationary neutral argon gas target, they found significantly lower
hardness ratios at collision energies comparable to those in their
EBIT \cite{2008PhRvA..78c2705A}. The lower hardness ratios in the beam
experiment are consistent with the predictions of CTMC theory.
\citet{2008PhRvA..78c2705A} consider numerous factors that could
affect CX either in the beam or in the trap and thus reconcile the
discrepancy, including the electric and magnetic fields in the trap,
the temperature of the ions in the trap, loss of photons from
metastable states in the extraction experiment, and possible polarized
and anisotropic emission in the extraction experiment, while emission
in the trap is unpolarized and isotropic. All of these factors are
considered unlikely by \citeauthor{2008PhRvA..78c2705A}, with the
exception of the possibility of polarization and anisotropy, which is
not yet well understood in the extraction experiment.

In this Letter we present high resolution x-ray spectra following CX
between highly ionized argon and phosphorus with molecular hydrogen.
Molecular hydrogen was chosen because of its relative simplicity as a
collision gas.  We compare the spectra and find that the hardness
ratio of H-like P is unexpectedly twice as large as that of H-like
Ar. The experiment was performed with bare P and Ar ions comixed in
the trap. Both ion species thus were exposed to identical experimental
conditions, e.g., the same neutral gases, magnetic and electric
fields, and trapping cycle length, and had the same temperature. The
fact that they produce starkly varying x-ray spectra shows that the
current understanding of charge exchange is insufficient and as of yet
unknown physics must be included in future x-ray production models.

Our measurements were performed using the SuperEBIT electron beam ion
trap at LLNL (\cite{2008CaJPh..86....1B} and references therein),
using the magnetic trapping mode \cite{1996RScI...67.3818B}. Neutral P
and Ar enter the trap and are subsequently ionized by the electron
beam. After the desired ionization balance has been achieved, the beam
is turned off, and EBIT operates as a Penning trap with the ions
confined in the radial direction by the $\sim 3$ T magnetic field of
the superconducting Helmholtz coils.

While the ions are magnetically trapped, H$_2$ is continuously
introduced through a ballistic gas injector at a much higher partial
pressure than Ar or P, and the only significant source of x-ray
emission is CX events between H$_2$ and the ions. After most of the
ions have filled the K shell through CX, the electric potential
trapping the ions along the axial direction is turned off and the trap
is dumped.  The trapping cycle was divided into a beam-on ionizing
phase (1.73s) and a beam-off CX phase (2.23s); we excluded the first 20
ms of the CX phase from our analysis to avoid contamination from the
beam-on phase.

Trace amounts of silicon, sulfur, and chlorine were also noted to
contribute indigenous ions in the trap.  Only the strongest
transitions of these trace species were detected, and in most cases
they do not blend with any features of Ar or P. The very weak emission
from He-like Cl ($\sim 25$ counts in He~$\alpha$) is blended with the
$n=8-10$ lines of He-like P ($\sim 6$ counts for all three lines); we
estimated the contribution of the blended lines (the resonance and
intercombination lines, w, x, and y) from the strength of the
forbidden line, z, and accounted for the blends in our modeling of the
high-$n$ He-like P lines.

The spectra were recorded using the XRS/EBIT x-ray calorimeter
instrument developed at NASA Goddard Space Flight Center
\cite{2004RScI...75.3772P, 2008CaJPh..86..231P}. The XRS/EBIT is a
high resolution ($\sim 6$ eV) nondispersive spectrometer that works
by accurately measuring the temperature change in absorbing pixels
operated at very low temperatures (60 mK). The XRS/EBIT has 28 pixels
with a 0.1-12 keV bandpass, of which 25 were operated.

The XRS/EBIT has four permanent aluminized polyimide windows that are
used for optical filtering and thermal isolation of the detector and
cooling system. An additional, much thicker aluminized polyimide
window was used to reduce the flux of soft x rays (mainly due to Ar
and P L-shell emission) at the detector.  We looked for evidence of
contamination on the filters by ice or hydrocarbons using
contemporaneous data from experiments using low-Z ions such as H-like
carbon and oxygen. We found no evidence for excess absorption over
that expected from the windows. The upper limit to the amount of
material accumulated on the windows is small enough that any
contaminant present would have a negligible effect on the transmission
above 2 keV.

\begin{figure*}
  \includegraphics[width=7in]{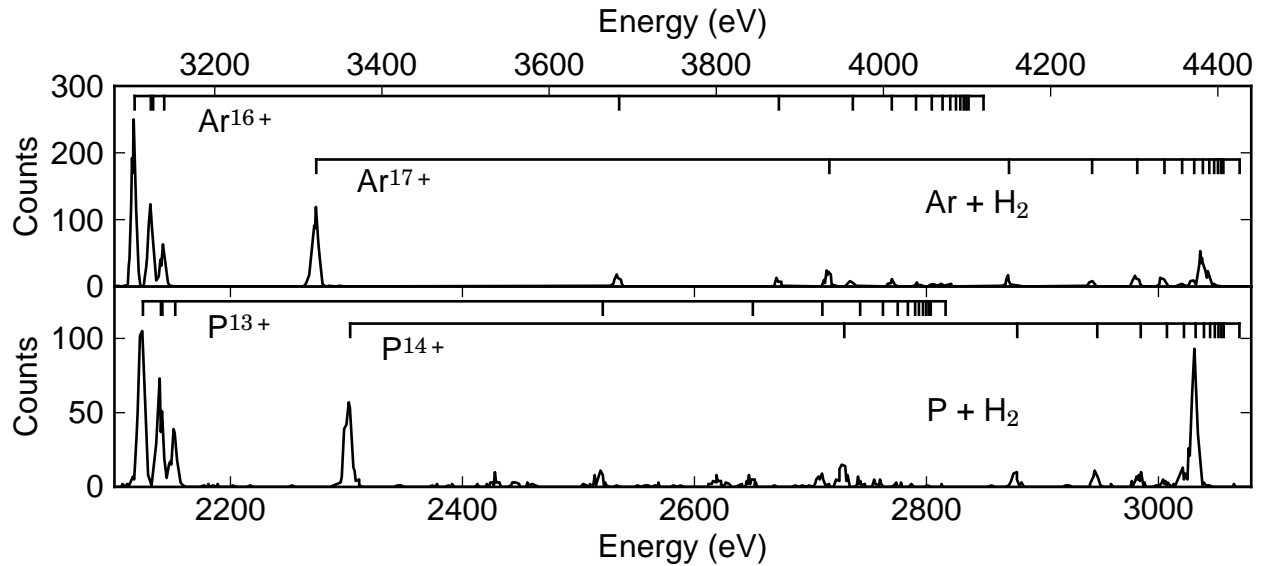}
  \caption{X-ray emission spectra of H-like and He-like argon and
    phosphorus following charge exchange with molecular hydrogen.  The
    rest energies of the series are indicated up to the $15p
    \rightarrow 1s$ transition, as well as the series limit.}
  \label{fig:spectra}
\end{figure*}

We present the concurrently observed CX spectra in Fig.
\ref{fig:spectra}. There is a striking difference between the two
hydrogenlike spectra; i.e., the relative intensity of the emission
from capture to the levels with the highest two principal quantum
numbers is 3 times larger for P than for Ar.  We measured the
number of counts in all lines by fitting them with Gaussians.  We
corrected the measured counts for the filter transmission and computed
the relative strengths of the lines in the H-like Lyman series
(Table~\ref{tab:Hseries}), as well as the He-like singlet series
($1s\, np\, ^1P_1 \rightarrow 1s^2\, ^1S_0$) and the He $\alpha$
complex (Table~\ref{tab:HeSeries}).  The quantum efficiency of the
detectors in the 2-5 keV range is nearly unity, so we do not correct
for it.

Tables~\ref{tab:Hseries} and \ref{tab:HeSeries} include measurements
of the hardness ratio, ${\cal H} \equiv F_{3+} / F_2$. In
Table~\ref{tab:HeSeries}, we also define a new quantity for the
He-like spectrum, ${\cal H}^\prime \equiv F_{3+} / F_w$ where $F_w$ is
the observed flux in the resonance line ($1s\, 2p\, ^1P_1 \rightarrow
1s^2\, ^1S_0$). For low-$Z$ systems such as Ar and P, where
transitions from $n > 2$ to ground are dominated by emission from
singlet states, this has the advantage of measuring the hardness ratio
only among the singlets, and does not contain information about the
triplet-to-singlet capture ratio. Also in Table~\ref{tab:HeSeries}, we
give both the ratios defined in \citet{1969MNRAS.145..241G}, ${\cal R}
\equiv F_z / (F_x + F_y)$ and ${\cal G} \equiv (F_z + F_y + F_x) /
F_w$, as well as an alternate quantity giving the ratio of
triplet-to-singlet captures, ${\cal G}^\prime \equiv (F_z + F_y + F_x)
/ (F_w + F_{3+})$.

\begin{table}
  \caption{Normalized H-like series line strengths and hardness
    ratio. Filter transmission corrected line strengths are normalized
    to 1000 for 1s-2p. Errors are statistical
    only.\label{tab:Hseries}}
  \begin{ruledtabular}
    \begin{tabular}{ccc}
             & Ar   & P \\
      $1s-2p$  & 1000 & 1000 \\
      $1s-3p$  & 176  & 248 \\
      $1s-4p$  & 78   & 133 \\
      $1s-5p$  & 53   & 151 \\
      $1s-6p$  & 102  & 111 \\
      $1s-7p$  & 103  & 50 \\
      $1s-8p$  & 20   & 152 \\
      $1s-9p$  & 66   & 1224 \\
      $1s-10p$ & 324  & ... \\
      $1s-11p$ & 117  & ... \\
      ${\cal H}$ & 1.04 $\pm$ 0.05 & 2.07 $\pm$ 0.12 \\
    \end{tabular}
  \end{ruledtabular}
\end{table}

\begin{table}
  \caption{Normalized He-like series line strengths and ratios. Filter
    transmission corrected line strengths are normalized to 1000 for
    $w$. Errors are statistical only. Definitions of the line
    ratios ${\mathcal R}$, ${\mathcal G}$, ${\mathcal G}^\prime$,
    ${\mathcal H}$, and ${\mathcal H}^\prime$ are given in the
    text. \label{tab:HeSeries}}
  \begin{ruledtabular}
    \begin{tabular}{ccc}
             & Ar & P \\
      $w$    & 1000 & 1000 \\ 
      $1s-3p$  & 296 & 233 \\ 
      $1s-4p$  & 186 & 145 \\ 
      $1s-5p$  & 120 & 190 \\ 
      $1s-6p$  & 134 & 75 \\ 
      $1s-7p$  & 55 & 36\\ 
      $1s-8p$  & 55 & 14 \\ 
      $1s-9p$  & 53 & 17 \\ 
      $1s-10p$ & 34 & ... \\ 
      ${\mathcal R}$       & 1.73 $\pm$ 0.07 & 2.27 $\pm$ 0.15 \\
      ${\mathcal G}$       & 6.0 $\pm$ 0.3   & 4.3 $\pm$ 0.3  \\
      ${\mathcal G}^\prime$ & 3.12 $\pm$ 0.13 & 2.55 $\pm$ 0.15 \\
      ${\mathcal H}$       & 0.133 $\pm$ 0.007 & 0.128 $\pm$ 0.010 \\
      ${\mathcal H}^\prime$ & 0.93 $\pm$ 0.07 & 0.68 $\pm$ 0.07 \\
    \end{tabular}
  \end{ruledtabular}
\end{table}

\begin{figure}[ht]
  \includegraphics[width=84mm]{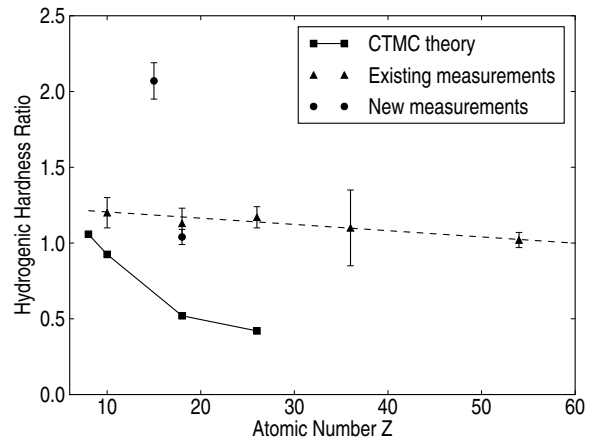}
  \caption{Measured and calculated values of the hardness ratio of
    H-like CX emission as a function of atomic number $Z$. The dashed
    line gives a linear fit to the previously measured hardness
    ratios.}
  \label{fig:HR}
\end{figure}

In Fig.~\ref{fig:HR} we plot the hardness ratios measured in
\citet{2000PhRvL..85.5090B} and \citet{2005ApJ...634..687W} as a
function of atomic number, along with the theoretical CTMC hardness
ratios computed in those papers, as well as the new measurements
reported in this Letter. All of the measurements were carried out
under conditions corresponding to low collision velocities. We
estimate the ion temperature in the present experiment to be $450 \pm
225$ eV \cite{1995RScI...66..303B, 2005ITPS...33.1763C}. For the same
ion temperature, the typical collision velocity of P is slightly
higher than for Ar by a factor of $\sqrt{M_{\mathrm{Ar}} /
  M_{\mathrm{P}}} = 1.14$.

The previous measurements are consistent with a linear, almost
constant, dependence of ${\mathcal H}$ on $Z$ (dashed line), although
the CTMC predictions are lower over the whole range in $Z$. The
hardness ratio we measure for Ar agrees very well with the earlier
measurements, which indicates that we were able to reproduce the
earlier experimental conditions and results. However, the new
measurement of the hardness ratio of P is a factor of 2
greater than any of the other measurements.

The large difference in hardness ratio we have measured for H-like
P and Ar is robust, since the spectra were obtained by
creating the bare ions of both elements {\it simultaneously in the
  same trap}, interacting with the same donor gas, and subject to the
same electric and magnetic fields and trap timing cycle. Although
there is some overall uncertainty in the ion temperature, and thus
collision velocity, the conditions for the two ions must be
comparable, since they interact with each other, are confined by the
same potential, and have similar charge and mass \cite{1991PhRvA..43.4873P}.

The measured He-like spectra of Ar and P are also significantly
different, as quantified by the line ratios reported in
Table~\ref{tab:HeSeries}. The triplet-to-singlet capture ratio ${\cal
  G}^\prime$ for He-like Ar is consistent with three, which is
expected from statistical considerations, but that of P is
not. Nonstatistical triplet ratios have been measured in charge
exchange between N$^{4+}$ and H or H$_2$ \cite{1998PhRvA..57..221B}.
Those measurements are in agreement with the fully quantal
close-coupled molecular orbital calculations of
\citet{1997JPhB...30.1013S}. However, in the case of our measurement
it is significant that captures involving two closely neighboring ion
species do not follow the same trend.

Dependence of angular momentum distributions of state-selective
capture cross sections $\sigma_{nl}$ on collision velocity has been
reported for single electron capture onto He-like species in multiple
experiments, with collision energies ranging down to 50 eV/amu
\cite{1985JPhB...18..737D, 1985JPhB...18.4763D,
  1990PhRvA..41.4800H}. These results could be qualitatively
understood in terms of classical considerations: first, the energy
levels of different angular momentum states determine their relative
degree of resonance with the reaction window, which becomes narrower
at low collision velocity; and second, the relative unimportance of
capture into $d$ and higher angular momentum states at low velocities
can be attributed to the low classical angular momentum of the
collisions.  Our finding that the capture cross section into the $9p$
state of H-like P is enhanced is surprising, since the
energy splitting of the $9l$ states is much smaller than the size of
the reaction window. For such highly charged ions, even the splitting
between $n$ levels is comparable to the width of the reaction
window. It is understandable that capture into high $l$ states
should be negligible at the low collision velocities in EBIT, but
there is no obvious reason for variations in the relative importance
of capture into $ns$ and $np$ in Ar and P.  It is likely that a fully
quantal molecular orbital calculation of capture cross sections will
be required to understand our results.

In summary, we have shown that two bare ion species under identical
experimental conditions produce strikingly different x-ray spectra
after charge exchange with H$_2$. The P spectrum is inconsistent with
all previous measurements, including those involving lower-Z bare
ions. Currently, there is no theory that suggests why this should be
the case, which means that there is no \textit{ab initio} way to
predict even the gross features of an x-ray emission spectrum not yet
measured in the laboratory.

\begin{acknowledgments}

 We would like to acknowledge the aid of D. Thorn and J. Clementson in
 data acquisition; M. F. Gu for calibration support; and E. Magee for
 technical support. MAL is supported by an appointment to the NASA
 Postdoctoral Program at Goddard Space Flight Center, administered by
 Oak Ridge Associated Universities through a contract with NASA. Part
 of this work was performed by Lawrence Livermore National Laboratory
 under the auspices of the US Department of Energy under Contract No.
 DE-AC52-07NA27344. The XRS/EBIT instrument was constructed and
 maintained with support from NASA.

\end{acknowledgments}

\bibliography{ebit}

\end{document}